\documentstyle[aps,epsfig]{revtex}
\begin{document}

\title{A Novel Non-Fermi Liquid \\ Ground State in One Dimension}
\author{Girish S. Setlur \\ Harish Chandra Research Institute \\
 Chhatnag Road, Jhusi, Allahabad, India 211 019.  } 

\maketitle

\begin{abstract}
 We propose a new non-Fermi liquid ground state in one dimension that is 
 not of the Luttinger type. It is the ground state of fermions
 interacting via a long range repulsive interaction in real space of the form
 $ V(x) = e^2/|x| $.
 We characterise it by computing the momentum
 distribution which is perfectly flat near the Fermi points signalling a 
 radical departure from Luttinger liquid theory(and Fermi liquid theory).
 The dispersion of elementary excitations is found to be sublinear but gapless
  of the form $ v\mbox{  } |q| \mbox{  }[ Log[2k_{F}/|q| ]^{\frac{1}{2}} $. 
 We also compute the conductance of a mesoscopic wire made of such objects
 and contrast it from the answer we obtain for a Luttinger wire.
\end{abstract}

\vspace{0.3in}

 Transport in interacting one dimensional wires has been a subject of intense
 study in the past few years starting with the works
 of Kane and Fisher\cite{Kane}. Since then a number of papers have appeared
 that discuss the generalisation to quantum wires with leads
 \cite{Sumathi1},
  multiple wire systems
 \cite{Sumathi2}, \cite{Chamon}. Recently Nazarov and Glazman
\cite{Nazarov}
 have developed an unusual perturbative scheme for understanding 
 resonant tunneling of interacting electrons in a one-dimensional wire.
 There do not seem to be many papers that discuss long-range interactions in
 one dimensional quantum wires. Kane, Balents and Fisher\cite{Balents} have considered the
 problem of Coulomb interactions which they have argued may be treated effectively
 using Luttinger liquid theory with the physics of Coulomb blockade\cite{Coulomb} added on. 
 However, in our earlier work\cite{Setlur} we showed that fermions interacting
 via a Coulomb interaction( or gauge interaction ) 
 namely $ v_{q} = const/q^2 $ is not characterisable using Luttinger liquid
 theory. Rather the ground state is a novel non-Fermi liquid which we
 characterised as a Wigner crystal. This is because in real space this
 pontetial is of the confining type
 namely $ V(x) \sim - |x| $. In this article we consider interactions between
 electrons of the type $ V(x) = e^2/|x| $, in other words the usual Coulomb
 interaction in
 three dimensions but the electrons being confined to move in one dimension. It is
 reasonable to
 suspect that real conductors in one dimension will have electrons interacting 
 via this type of interaction. It is repulsive unlike in the gauge interaction
 which is confining and attractive. Short-range interactions are due to
 screening and screening is a dynamical phenomenon. The `screening-first'
 approaches have been rightly critcised by Castro-Neto and
 Fradkin\cite{Fradkin} but the community continues to employ the Luttinger
 liquid paradigm. In this article we investigate what changes are 
 brought about to the computations of conductance, momentum distribution 
 and so on by this kind of long-range interaction. Kane, Balents and Fisher
 have found that the high temperature regime is characterized
 by Luttinger liquid physics but the low temperature regime is novel.
 We write the model for interacting electrons in a wire of 
 length $ L $ in the sea-boson language\cite{Setlur}. 
\[
H = \sum_{k,q, \sigma, \sigma^{'} }
 \frac{k.q}{m} A^{\dagger}_{ k \sigma }(q\sigma^{'})
A_{ k \sigma }(q\sigma^{'})
\]
\[
+ \sum_{ q \neq 0 } \frac{ v_{q} }{2L}
\sum_{k \neq k^{'}}\sum_{ \sigma, \sigma^{'} }
[A_{ k \sigma }(-q\sigma) + A^{\dagger}_{ k \sigma }(q\sigma)]
[A_{ k^{'} \sigma^{'} }(q\sigma^{'}) +
 A^{\dagger}_{ k^{'} \sigma^{'} }(-q\sigma^{'})]
\]
\begin{equation}
-  \sum_{ q \neq 0 } 
\sum_{k \neq k^{'}}\frac{ v_{k-k^{'}} }{2L} 
\sum_{ \sigma, \sigma^{'} }
[A_{ k \sigma }(-q\sigma^{'}) + A^{\dagger}_{ k \sigma^{'} }(q\sigma)]
[A_{ k^{'} \sigma^{'} }(q\sigma) +
 A^{\dagger}_{ k^{'} \sigma }(-q\sigma^{'})]
\end{equation}
 Here $ k = n \pi/L $,
 $ q = 2 n \pi/L $ and $ n = 0, \pm 1, \pm 2, ... $.
 The potential $ v_{q} = (2 e^2) \mbox{        } Log[ \frac{ 2 k_{F} }{ |q| }
 ]  $. Now we may simplify the above
 hamiltonian retaining only the most singular parts. The part that does not
 flip spins may be written down as shown.
\[
H_{c} = \sum_{k,q, \sigma }
 \frac{k.q}{m} A^{\dagger}_{ k \sigma }(q\sigma)
A_{ k \sigma }(q\sigma)
\]
\begin{equation}
- \sum_{ q \neq 0 } \frac{ e^2 \mbox{        } Log[ \frac{ |q| }{ (2k_{F}) } ] }{L}
\sum_{ k,k^{'},\sigma,\sigma^{'} }
[A_{ k \sigma }(-q\sigma) + A^{\dagger}_{ k \sigma }(q\sigma)]
[A_{ k^{'} \sigma^{'} }(q\sigma^{'}) +
 A^{\dagger}_{ k^{'} \sigma^{'} }(-q\sigma^{'})]
\end{equation}
The singular part of the spin flipping hamiltonian is,
\[
H_{s} = \sum_{ q \neq 0 } 
\sum_{k \neq k^{'}, \sigma }\frac{ 2 e^2 \mbox{        } Log[ \frac{ |k-k^{'}| }{2k_{F}} ] }{L} 
 A^{\dagger}_{ k {\bar{\sigma}} }(q\sigma)
A_{ k^{'} {\bar{\sigma}} }(q\sigma)
\]
\begin{equation}
+ \sum_{ q \neq 0 } 
\sum_{k,k^{'}, \sigma }\frac{ e^2\mbox{        }
 Log[ \frac{ |k-k^{'}| }{ (2k_{F}) } ] }{L} 
[A_{ k \sigma }(-q {\bar{\sigma}}) A_{ k^{'} {\bar{\sigma}} }(q\sigma)
+ A^{\dagger}_{ k {\bar{\sigma}} }(q\sigma)A^{\dagger}_{ k^{'} \sigma }(-q {\bar{\sigma}} )]
\end{equation}
Thus we already see evidence of spin-charge separation since the two
hamiltonians look very different. First we make some remarks about the 
spin-flip part of the hamiltonian. The first term involves a singular
logarithm that is large and negative near $ k = k^{'} $.
  The $ k \neq k^{'} $ restriction comes about from the fact that
 we are only considering correlation part of the hamiltonian.  
 Thus there is no macroscopic occupation of the $ k = k^{'} $ state
 since this contribution is absent. Furthermore, we argue that 
 in the high density limit the sencond term is small since $ |k-k^{'}| $
 is very close to $ 2k_{F} $. Thus in these circumstances it is legitimate to
 set $ < A^{\dagger}_{ k {\bar{\sigma}} }(q\sigma)
A_{ k {\bar{\sigma}} }(q\sigma) > = 0 $. The boson occupation is
 needed to compute the momentum distribution of the electrons.
 To compute the current-current correlation function we only need
 the operators that appear in $ H_{c} $. First we focus on the momentum
 distribution.
The boson occupation reads as follows.
\begin{equation}
\left< A^{\dagger}_{ k \sigma }(q\sigma)A_{ k \sigma }(q\sigma) \right>
 =  \frac{ 2 \pi k_{F} }{L} \frac{ \Lambda_{k}(-q) }
{ 2 \omega_{c}(q) ( \omega_{c}(q) + \frac{k.q}{m} )^2 \left( \frac{m^3}{q^4}
 \right) [ Cosh[\lambda(q)] - 1 ] }
\end{equation}
Here $ \lambda(q) = (\pi q/m) (1/v_{q}) $ and the dispersion of collective
modes may be obtained by computing the zeros of the RPA dielectric function.,
This may be simplified by expanding in the small $ q $ limit.
\begin{equation}
\omega_{c}(q) \approx \left( \frac{ 4 e^2 }{ \pi v_{F}}
\right)^{\frac{1}{2}}
 \left( Log\left[ \frac{ 2k_{F}
    }{ |q| } \right] \right)^{\frac{1}{2}} \mbox{      } v_{F} |q| 
\end{equation}
Therefore we find that even though the dispersion is gapless, it is not linear.
The momentum distribution is obtained in the usual way from the sea-boson
equations\cite{Setlur}.
\begin{equation}
<n_{k\sigma}> = \frac{1}{2} \left( 1 + e^{ -2 \sum_{q} <A^{\dagger}_{
    k+q/2 \sigma }(q\sigma) A_{ k+q/2 \sigma }(q\sigma) > }
    \right)n_{F}(k)
+  \frac{1}{2} \left( 1 - e^{ -2 \sum_{q} <A^{\dagger}_{
    k-q/2 \sigma }(q\sigma) A_{ k-q/2 \sigma }(q\sigma) > }
    \right)(1-n_{F}(k))
\end{equation}
 It is not possible to evaluate this analytically. Hence we plot
 this using $ Mathematica^{TM} $. This is depicted in Fig.1 below.
\begin{figure}
       \begin{center}
       \mbox{\psfig{figure=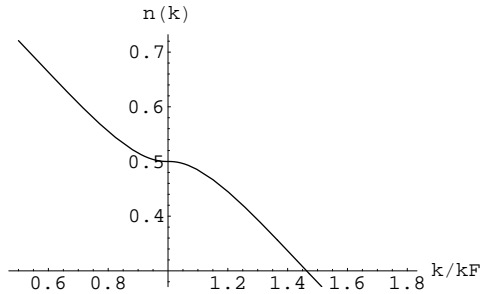,width=2.5in}}
       \caption{Momentum Distribution}
       \label{fig:Fig.1}
       \end{center}
\end{figure} 
 The flat nature of the momentum distribution near $ k_{F} $ shows a radical
 departure from Luttinger liquid theory. This should be accesible to tunneling
 measurements of the spectral function.

 To compute the conductance we have to use the Kubo formula. However, in
 mesoscopic physics
 we may not take the long-wavelength limit first since the system sizes are
 small of the order of a hundred atomic spacings. In this case we have to use
 the formula for the finite $ q $ version of the current-current correlator. 
 From the text by Rao\cite{Sumathi} the current flowing through a point $ x $ 
 in response to an electric field $ E $ may be written as follows,
\begin{equation}
 I(x) = \int^{L}_{0} dx^{'} \mbox{           }
\int^{ \infty }_{ -\infty } \frac{ d \omega }{ 2 \pi }
e^{ -i \omega t } \mbox{       }
\sigma_{ \omega }(x,x^{'}) \mbox{          }E_{ \omega }(x^{'})
\end{equation}

\begin{equation}
\sigma_{ \omega }(x,x^{'}) = -\frac{ i e^2 }{ m^2 \omega }
\int^{ -i \beta }_{0} dt \mbox{       }\left< j(x,t)j(x^{'},0)\right> \mbox{       }e^{ \omega t}
\end{equation}
 Since $ q $ is not vanishingly small, we may retain the linear terms
 in the sea-bosons and we have the following expression for the current.
\begin{equation}
j(x,t) = \frac{1}{L}\sum_{k,q\sigma} k \mbox{         }
 [A_{k\sigma}(-q\sigma,t) + A^{\dagger}_{k\sigma}(q\sigma,t)] \mbox{         }
e^{-iqx}
\end{equation}
This means we may use the simple form of current algebra to deduce the
current-current correlation function(in the case $ v_{q} \neq 0 $).
\begin{equation}
\left< j_{-q\sigma}(t)
j_{q\sigma}(0) \right> \approx \left( \rho^{0} \frac{ v_{q} }{2} \right)  \mbox{       }
\left( \frac{ q^2 N }{2} \right) \mbox{    }
 e^{ -i\omega_{c}(q) t } 
\mbox{          }
\frac{1}{ 2 \omega_{c}(q) }
\end{equation}
Thus we may evaluate the a.c. conductance as follows.
\begin{equation}
Re[ \sigma ]_{ \omega }(x,x^{'}) = -\frac{ e^2 }{ \omega }
\mbox{       }
\frac{1}{2 \pi}\int^{ \infty }_{ -\infty } dq \mbox{      }
 e^{ -i q(x-x^{'}) }
\mbox{         }
  (2 e^2) Log \left[ \frac{ 2 k_{F} }{ |q| } \right]  \mbox{       }
\left( \frac{ q^2 v^2_{F} }{ \pi^2 } \right) \mbox{    }
\mbox{          }
 \mbox{       }\frac{ 1 }
{ ( \omega^2 + \lambda^2 q^2\mbox{          }
  Log \left[ \frac{ 2 k_{F} }{ |q| } \right] ) }
\end{equation}
To evaluate this we have to make a branch-cut to pole approximation. To this
end we observe that the dominant contribution to this integral is coming from
not from the slowly varying logarithms but from the oscillating exponential
 times $ q^2 $. We evaluate the $ q $ at which $ cos[q(x-x^{'})] \mbox{
 }q^2 $  is a maximum. In other words, $ 2 = q (x-x^{'})\mbox{
 }tan[q(x-x^{'})] $. Or, $ |q_{opt}| \approx 1/|x-x^{'}| $.
\begin{equation}
Re[ \sigma ]_{ \omega }(x,x^{'}) = -\frac{ e^2 }{ \omega }
\mbox{       }
\frac{1}{2 \pi}\int^{ \infty }_{ -\infty } dq \mbox{      }
 e^{ -i q(x-x^{'}) }
\mbox{         }
  (2 e^2) Log \left[ 2 k_{F} |x-x^{'}| \right]  \mbox{       }
\left( \frac{ q^2 v^2_{F} }{ \pi^2 } \right) \mbox{    }
\mbox{          }
 \mbox{       }\frac{ 1 }
{ ( \omega^2 + \lambda^2 q^2\mbox{          }
  Log \left[ 2 k_{F} |x-x^{'}|\right] ) }
\end{equation}
Thus for large enough separations we have a pole. Therefore we may write, 
\begin{equation}
Re[ \sigma ]_{ \omega }(x,x^{'}) = 
 e^{ - \omega \frac{ |x-x^{'}| }{ \lambda (Log[2k_{F} |x-x^{'}|])^{\frac{1}{2}} }  }
\mbox{         } \mbox{       }
\left( \frac{ e^4 v^2_{F} }{ \pi^2 \lambda^3 } \right) \mbox{    }
\mbox{          }
 \mbox{       }\frac{ 1 }
{ (Log[2k_{F} |x-x^{'}|])^{\frac{1}{2}} }
\end{equation}
 where $ \lambda = ( 4 e^2 v_{F}/\pi )^{\frac{1}{2}} $. Therefore in the static
 limit, the conductance is finite and decays slowly for large separations.
 Using an STM it may be possible to measure this conductance and demonstrate its
 slow decay as a function of tip separation. It is instructive to contrast
 this  with the result for a Luttinger wire, namely $ \sigma( \omega = 0 ) = K e^2/(2\pi) $,
 independent of the separation\cite{Sumathi}. This can be understood intuitively by
 observing that in the case of long-range interactions, the electron that is
 carrying the current will feel resistance from all the electrons in the wire
 and not just from the ones immediately in its vicinity as is the case in a
 Luttinger wire. Thus the length of
 the wire is likely to determine the magnitude of the conductance. 

 It is a pleasure to thank Sumathi Rao for useful discussions. This work was
 supported by the Harish Chandra Research Institute.

\end{document}